\documentclass[12pt]{article}
\usepackage[dvips]{graphicx}
\usepackage{amssymb}
\usepackage{amsmath}

\def\beq{\begin{equation}}\def\eeq{\end{equation}}
\def\bea{\begin{eqnarray}}\def\eea{\end{eqnarray}}

\begin{document}

\title{How Mensky's continuous measurement can emerge from GRW on larger time scales}
\author{Roman Sverdlov, Department of Mathematics, University of New Mexico} 

\date{August 10, 2016}
\maketitle

\begin{abstract}
In this paper we will show how "weighted path integral" proposed by Mensky and Kent can emerge out of Ghirardi Rimini Weber (GRW) model.   \end{abstract}

\subsection*{Introduction}

Mensky (\cite{Mensky1} and \cite{Mensky2}) and Kent (\cite{Kent}) has proposed a model of continuous measurement. The probability that the particle follows a trajectory $x_{cl} (t)$ is given by a weighted path integral, 
\beq \int [{\cal D} x] e^{iS (x)} w (x, x_{cl}) \eeq
where
\beq S (x) = \int dt \; \bigg( \frac{m \dot{x}^2 (t)}{2} - V(x(t)) \bigg) \eeq
and $w (x,x_{cl})$ is a "weight factor", that satisfies the following properties

1. If $x_1$ and $x_2$ are classically distinguishable (or, equivalently, if we never have interference between the two) then $w (x_1, x_2) \approx 0$

2. If $x_1$ and $x_2$ are both $N$-particle configurations that agree regarding the location of $N-n$ particles up to the distance $l$ and, furthermore, they agree regarding the location of $n$ particle up to distance $L$ (as a toy example, lets say $l$ is one micron, $L$ is one kilometer, $n=10$ and $N=10^6$) then $w (x_1, x_2) \approx 1$.

The consequence of 1 is the existence of classical paths that obey the principle of least action; the consequence of 2 is few particle quantum mechanics. One example of $w$ that obeys the above properties is 
\beq w (x_1, x_2) = \exp \bigg( - \frac{\alpha}{2} \sum_{k=1}^N \vert \vec{x}_2^k - \vec{x}_1^k \vert^2 \bigg) \label{Gaussian}\eeq 
where 
\beq \vert \vec{x}_2^k - \vec{x}_1^k \vert^2 = (x_2^{k1} - x_1^{k1})^2 + (x_2^{k2} - x_1^{k2})^2 + (x_2^{k3} - x_1^{k3})^2 \eeq

In this paper, we would like to show how the above can emerge as a large scale approximation out of Ghirardi Rimini Weber (GRW) model (\cite{GRW1}, \cite{GRW2}). According to the model, the wave function evolves according to Schr\"odinger's equation, 
\beq i \frac{\partial \psi}{\partial t} = - \frac{1}{2m} \Delta \psi + V \psi \label{Schrodinger} \eeq
which is being interrupted by "hits" at times $\{ \cdots, t_{-2}, t_{-1}, t_0, t_1, t_2, \cdots \}$ which are the discrete changes to $\psi$ given by  
\beq \psi (x, t_k^+) = N (x_k, t_k) \psi (x, t_k^-) e^{- \frac{\alpha}{2} \vert x-x_k \vert^2} \label{hit} \eeq 
where $N (x_k,t_k)$ is a normalization constant:
\beq N (x_k, t_k) = \bigg( \frac{\int \vert \psi (x, t_k^-) \vert^2}{\vert \int  \vert \psi (x, t_k^-) \vert^2 e^{- \alpha \vert x - x_k \vert^2} \vert} \bigg)^{1/2} \eeq 
The point $x_k$ is randomly selected, by using "biased coin" with probability density being 
\beq \rho (x_k =x) = \frac{1}{N^2 (x,t_k)} \label{HitDensity}\eeq 
where by $x$ and $x_k$ we really mean configurations of particles:
\beq x = (\vec{x}^1, \cdots, \vec{x}^N) \; , \; x_k =  (\vec{x}_k^1, \cdots, \vec{x}_k^N) \eeq
and $\vec{x}_k^l$, in turn, lives in $\mathbb{R}^3$:
\beq \vec{x}_k^l = (x_k^{l1}, x_k^{l2}, x_k^{l3}) \eeq

The way we propose to derive continuous measurement model, described earlier, from GRW, is as follows. First of all, we re-think the meaning we attribute to $x_k$. Typically, within the context of GRW model, it is assumed that the reality is either $\psi (x)$ itself, or a mass density derived from it, while $x_k$ is just an outside factor that modifies its evolution. In this paper we will change things around and say that reality is $x_k$, \emph{as opposed to} aforementioned entities. In particular, $x_k$ describes a location of point particle; or, in Bohmian language, $x_k$ is a "position beable". However, instead of following a continuous trajectory, the "particle" appears for the infinitesimal period of time, then disappears, then a while later appears at a different location for infinitesimal time, then disappears again, and so forth. Now, on a time scale that is significantly larger than the time interval between appearances, it would seem as it it follows certain trajectory. That trajectory would not be continuous: indeed the location of the next point will be independent of the location of the previous one. Still, if we call it a trajectory, said trajectory will have a probability: namely the product of probabilities of each appearance. And we will derive that, indeed, said probability will approximate the one proposed by Mensky (\cite{Mensky1} and \cite{Mensky2}) and Kent (\cite{Kent}) with weight function given by Eq \ref{Gaussian}. 

This has some vague similarity with more conventional way of identifying the reality with mass density. In both cases, the actual dynamics governs wave function $\psi$, which lives in configuration space, and that wavefuction produces a ``shadow" that lives in ordinary $\mathbb{R}^3$, and that ``shadow" is identified with reality. In conventional case, the ``shadow" is mass density, 
in our case the ``shadow" is the position of hits. Mass density lives in $\mathbb{R}^3$ rather than $\mathbb{R}^{3n}$ because the way it is derived from $\psi$ is onto but not one to one: for example, the state $\vert \psi_1 \rangle \otimes \vert \psi_2 \rangle$ might produce the same mass density as $\vert \psi_1 \rangle \oplus \vert \psi_2 \rangle$ (see Sec 3.2 of \cite{Referree1}). Likewise the location of hits can be seen as a single point in $\mathbb{R}^{3n}$ which, in turn, is in bijective correspondence with a set of $n$ points in $\mathbb{R}^3$ which, in turn, can be represented as a sum of $n$ different $\delta$-functions in $\mathbb{R}^3$ space\footnote{Some versions of GRW model subject just one particle to a hit instead of the entire configuration of $n$ particles. In those cases, we would have a single $\delta$-function in $\mathbb{R}^3$ instead of superposition of $n$ such $\delta$-function; but this won't affect the ultimate conclusion that the reality is in $\mathbb{R}^3$. In any case, as far as this paper is concerned, our assumption is that we subject a configuration of $n$-particles to simultaneous hit}. The difference, however, is that if we know the wave function at $t=t_0$ we know the mass density at $t=t_0$. With the hit its not that simple: our knowledge of wave function at $t=t_0^-$ will only tell us the location of a hit at $t=t_0$ with a certain probability; although the comparison of the wave function at $t=t_0^-$ and $t=t_0^+$ will determine the hit with certainty. The other difference is that mass density changes continuously with $\psi$ whereas hits only occur during discretely-separated from each other infinitesimal time intervals. One could expect, however, that on some really rough scale the two might approximate each other over time: after all, the effect of hits includes concentrating mass density within a region of a certain size around where the hit occurred. However, what makes it preferable to identify the reality with hits themselves is that, as the next section will show, the correspondence to Mensky and Kent path integral is more direct in this way. If we identify reality with mass density, the correspondence would be less direct, which means that the probability of mass density trajectory might include some correction terms (either probabilistic and/or deterministic) investigation of which is beyond the scope of this paper. 

Even though we have been insisting how we identify the reality with $x_k$ rather than $\psi$, we are not restricted to this viewpoint. If we want to go back to $\psi$-based reality, we can go back to the explanation of GRW model the way it is originally given and, instead, change the explanation of Mensky and Kent. In particular, instead of viewing $x_{cl}$ as reality, we view $x_{cl}$ (which is still selected as prescribed by Mesky and Kent) as just an intermediate step which affects subsequent evolution of $\psi$, which, in turn, produces a mass density, distinct from $x_{cl}$; we are then free to choose to identify reality with the mass density and $x_{cl}$ as merely an intermediate step of producing it.  In particular, the evolution of $\psi$ is given by
\beq \psi (x_0, t=T_2) = \int_{x(T_2)=x_0} [{\cal D} x] \psi (x(T_1); t=T_1) e^{iS (x)} w (x, x_{cl}) \eeq
and $x_{cl}$ figures in $w (x,x_{cl})$ on the right hand side. It is easy to see that, if we replace $x_{cl}$ with "hits" at locations $x_{cl} (t_k)$ at times $t=t_k$, and make sure $t_k$ are selected densely enough, then the modification to $\psi$ described above (and consequent modification to mass density) will approximate the one given by GRW model. The only difference between two viewpoints is that, if we assume continuity of $x_{cl}$, then two subsequent hits in continuous measurement case will be a lot closer than they would be in GRW model. But, if we abandon the assumption that $x_{cl}$ is continuous, then the two views will match more closely. 

\subsection*{Calculation}

Suppose we know the wave function at $t_M^+$, and we would like to ask ourselves what is the probability that we will ''hit'' the ''afore-given'' values of $\{\vec{x}_{M+1}, \cdots, \vec{x}_N \}$. Suppose we have already succeeded ''hitting'' $\{\vec{x}_{M+1}, \cdots, \vec{x}_{k-1} \}$. Then the probability of ''hitting'' $\vec{x}_k$ is
\beq f_k (\vec{x}_k) = \bigg(\frac{\alpha}{\pi}\bigg)^{3/2} \int d^3 x' e^{- \alpha \vert \vec{x}'-\vec{x}_k \vert^2} \vert \psi (\vec{x}',t^-) \vert^2 \label{f} \eeq
As we recall, $\psi$ evolved according to Equations \ref{Schrodinger} and \ref{hit}. If we now rewrite Eq \ref{Schrodinger} in terms of path integral, we obtain
\beq \psi (\vec{x},t_{j+1}^-) = A_{j+1} \int_{\vec{x}'(t_{j+1}) = \vec{x}} [{\cal D} x'] \psi (\vec{x}', t_j^+) \exp \bigg(i \int_{t_j}^{t_{j+1}} dt {\cal L} (\vec{x}', t') \bigg) \label{SegmentPath}\eeq
\beq \psi (t_{j+1}^+, \vec{x}) = B_{j+1} e^{\frac{\alpha}{2} \vert \vec{x} - \vec{x}_{j+1} \vert^2} \psi (t_{j+1}^-, \vec{x}) \label{hitPath} \eeq
By substituting Eq \ref{SegmentPath} into Eq \ref{hitPath} we obtain
\beq \psi (\vec{x},t_{j+1}^+) = A_{j+1} B_{j+1} \int_{\vec{x}'(t_{j+1}) = \vec{x}} [{\cal D} x'] \psi (\vec{x}', t_j^+) \exp \bigg(-\frac{\alpha}{2} \vert \vec{x} - \vec{x}_{j+1} \vert^2 + i \int_{t_j}^{t_{j+1}} dt \; {\cal L} (\vec{x}', t') \bigg) \label{SegmentPath2}\eeq
By substituting $\psi (\vec{x}, t^+_{j+1})$ into $\psi (\vec{x}, t^+_j)$ we can now compute $\psi (\vec{x}, t^+_{j+2})$ as follows:
\beq \psi (\vec{x},t_{j+2}^+) = A_{j+2} B_{j+2}\int_{\vec{x}'(t_{j+2}) = \vec{x}} [{\cal D} x'] \psi (\vec{x}', t_{j+1}^+) \exp \bigg(-\frac{\alpha}{2} \vert \vec{x} - \vec{x}_{j+2} \vert^2 + i \int_{t_{j+1}}^{t_{j+2}} dt \; {\cal L} (\vec{x}', t') \bigg) = \nonumber \eeq
\beq = A_{j+1} A_{j+2} B_{j+1} B_{j+2} \int_{\vec{x}'(t_{j+2}^-) = \vec{x}} [{\cal D} x'] \bigg[ \exp \bigg(- \frac{\alpha}{2} \vert \vec{x} - \vec{x}_{j+2} \vert^2 +i \int_{t_{j+1}}^{t_{j+2}} dt \; {\cal L} (\vec{x}', t') \bigg) \times \eeq
\beq \times \int_{\vec{x}''(t_{j+1}) = \vec{x}' (t_{j+1})} [{\cal D} x''] \psi (\vec{x}'', t_j^+) \exp \bigg(-\frac{\alpha}{2} \vert \vec{x}' (t_{j+1})- \vec{x}_{j+1} \vert^2 + i \int_{t_j}^{t_{j+1}} dt' \; {\cal L} (\vec{x}'', t') \bigg) \bigg] \nonumber \eeq
In the above expression, $\vec{x}'$ runs through $t_{j+1} \leq t < t_{j+2}$ and $\vec{x}''$ runs through $t_j \leq t < t_{j+1}$. We can therefore combine them into a single $\vec{x}'$, running through $t_j \leq t < t_{j+2}$. In this case the condition of the second integral, $\vec{x}'' (t_{j+1}^-) = \vec{x}'$, can be dropped, since it is a simple consequence of continuity of $\vec{x}'$. The two integrals combine into one to produce
\beq \psi (\vec{x},t_{j+2}^+) =A_{j+1} A_{j+2} B_{j+1} B_{j+2} \times \eeq
\beq \times \int_{\vec{x}'(t_{j+2}) = \vec{x}} [{\cal D} x'] \psi (\vec{x}', t_j^+) \exp \bigg(-\frac{\alpha}{2} \vert \vec{x}' - \vec{x}_{j+1} \vert^2 - \frac{\alpha}{2} \vert \vec{x} - \vec{x}_{j+2} \vert^2 +i \int_{t_j^+}^{t_{j+2}^-} dt' \; {\cal L} (\vec{x}', t') \bigg) \nonumber \eeq
By induction, one can see that 
\beq \psi (\vec{x}, t_k^-) = \bigg(\prod_{i=M+1}^k A_i \bigg) \bigg(\prod_{j=M+1}^{k-1} B_j \bigg) \times \eeq
\beq \times \int_{\vec{x}' (t_k)= \vec{x}} [{\cal D} x'] \psi (\vec{x}', t_M^+) \exp \bigg(-\frac{\alpha}{2} \sum_{j=M+1}^{k-1} \vert \vec{x} - \vec{x}_j \vert^2 + i \int_{t_M}^{t_k} dt' \; {\cal L} (\vec{x}', t') \bigg) \nonumber \eeq
Since $A_j$-s and $B_j$-s were selected in such a way that $\psi$ was normalized properly at any given step, we know by induction that their product will produce correct normalization at the end. This allows us to ''know'' the product in question without explicitly computing each of its ingredients. Therefore, we obtain
\beq \psi_{M,k-1} (\vec{x},t) = \eeq
\beq = \frac{ \int_{\vec{x}' (t_k)= \vec{x}} [{\cal D} x'] \psi (\vec{x}', t_M^+) \exp \bigg(-\frac{\alpha}{2} \sum_{j=M+1}^{k-1} \vert \vec{x} - \vec{x}_j \vert^2 + i \int_{t_M}^{t_k} dt' \; {\cal L} (\vec{x}', t') \bigg)}{\bigg( \int d^3 y \bigg\vert \int_{\vec{y}' (t_k)= \vec{y}} [{\cal D} y'] \psi (\vec{y}', t_M^+) \exp \bigg(-\frac{\alpha}{2} \sum_{j=M+1}^{k-1} \vert \vec{y} - \vec{x}_j \vert^2 + i \int_{t_M}^{t_k} dt' \; {\cal L} (\vec{y}', t') \bigg) \bigg\vert^2 \bigg)^{1/2}} \nonumber \eeq
By substituting this into Eq \ref{f}, we find that
\beq \bigg( \frac{\pi}{\alpha} \bigg)^{3/2} f_k (\vec{x}_k) = \eeq
\beq = \frac{ \int d^3 x e^{- \alpha \vert \vec{x} - \vec{x}_k \vert^2} \bigg\vert \int_{\vec{x}' (t_k)= \vec{x}} [{\cal D} x'] \psi (\vec{x}', t_M^+) \exp \bigg(-\frac{\alpha}{2} \sum_{j=M+1}^{k-1} \vert \vec{x} - \vec{x}_j \vert^2 + i \int_{t_M}^{t_k} dt' \; {\cal L} (\vec{x}', t') \bigg) \bigg\vert^2}{\int d^3 y \bigg\vert \int_{\vec{y}' (t_k)= \vec{y}} [{\cal D} y'] \psi (\vec{y}', t_M^+) \exp \bigg(-\frac{\alpha}{2} \sum_{j=M+1}^{k-1} \vert \vec{y} - \vec{x}_j \vert^2 + i \int_{t_M}^{t_k} dt' \; {\cal L} (\vec{y}', t') \bigg) \bigg\vert^2} \nonumber \eeq
By absorbing $e^{- \alpha \vert \vec{x} - \vec{x}_k \vert^2}$ into the summation of $\vert \vec{x} - \vec{x}_j \vert^2$, we obtain
\beq \bigg( \frac{\pi}{\alpha} \bigg)^{3/2} f_k (\vec{x}_k) = \label{Ratio1}\eeq
\beq = \frac{ \int d^3 x \bigg\vert \int_{\vec{x}' (t_k)= \vec{x}} [{\cal D} x'] \psi (\vec{x}', t_M^+) \exp \bigg(-\frac{\alpha}{2} \sum_{j=M+1}^k \vert \vec{x} - \vec{x}_j \vert^2 + i \int_{t_M}^{t_k} dt' \; {\cal L} (\vec{x}', t') \bigg) \bigg\vert^2}{\int d^3 y \bigg\vert \int_{\vec{y}' (t_k)= \vec{y}} [{\cal D} y'] \psi (\vec{y}', t_M^+) \exp \bigg(-\frac{\alpha}{2} \sum_{j=M+1}^{k-1} \vert \vec{y} - \vec{x}_j \vert^2 + i \int_{t_M}^{t_k} dt' \; {\cal L} (\vec{y}', t') \bigg) \bigg\vert^2} \nonumber \eeq
Now, from Eq \ref{SegmentPath}, we know that 
\beq \psi (\vec{y},t_k^-) = A_k \int_{\vec{x}'(t_k) = \vec{y}} [{\cal D} y'] \psi (\vec{y}', t_{k-1}^+) \exp \bigg(i \int_{t_{k-1}}^{t_k} dt {\cal L} (\vec{y}', t') \bigg) \eeq
This implies that if we are going to replace ''integral from $t_M$ to $t_k$'' with ''integral from $t_M$ to $t_{k-1}$'' in the denominator of Eq \ref{Ratio1}, we will change the ratio by the factor of $A_k$:
\beq A_k \bigg( \frac{\pi}{\alpha} \bigg)^{3/2} f_k (\vec{x}_k) = \label{Ratio2}\eeq
\beq = \frac{ \int d^3 x \bigg\vert \int_{\vec{x}' (t_k)= \vec{x}} [{\cal D} x'] \psi (\vec{x}', t_M^+) \exp \bigg(-\frac{\alpha}{2} \sum_{j=M+1}^k \vert \vec{x} - \vec{x}_j \vert^2 + i \int_{t_M}^{t_k} dt' \; {\cal L} (\vec{x}', t') \bigg) \bigg\vert^2}{\int d^3 y \bigg\vert \int_{\vec{y}' (t_k)= \vec{y}} [{\cal D} y'] \psi (\vec{y}', t_M^+) \exp \bigg(-\frac{\alpha}{2} \sum_{j=M+1}^{k-1} \vert \vec{y} - \vec{x}_j \vert^2 + i \int_{t_M}^{t_{k-1}} dt' \; {\cal L} (\vec{y}', t') \bigg) \bigg\vert^2} \nonumber \eeq
We can rewrite it as 
\beq A_k \bigg( \frac{\pi}{\alpha} \bigg)^{3/2} f_k (\vec{x}_k) = \frac{I_{M,k}}{I_{M,k-1}} \eeq
where
\beq I_{M,k}= \int d^3 x \bigg\vert \int_{\vec{x}' (t_k)= \vec{x}} [{\cal D} x'] \psi (\vec{x}', t_M^+) \exp \bigg(-\frac{\alpha}{2} \sum_{j=M+1}^k \vert \vec{x} - \vec{x}_j \vert^2 + i \int_{t_M}^{t_k} dt' \; {\cal L} (\vec{x}', t') \bigg) \bigg\vert^2 \eeq
Now that we have computed the probability of any individual hit, we can go back to our original goal of computing the probability of sequence of hits, $(\vec{x}_{M+1}, t_{M+1}), \cdots, (\vec{x}_N, t_N))$. The latter is 
\beq f_{M+1} (\vec{x}_{M+1}), \cdots, f_N (\vec{x}_N) = \eeq
\beq = \bigg(\prod_{k=M+1}^N A_k \bigg) \bigg(\frac{\alpha}{\pi}\bigg)^{\frac{3}{2} (N-M)} \frac{I_{M,M+1}}{I_{M,M}} \cdots \frac{I_{M,N}}{I_{M,N-1}} = \bigg(\prod_{k=M+1}^N A_k \bigg) \bigg(\frac{\alpha}{\pi}\bigg)^{\frac{3}{2} (N-M)} \frac{I_{M,N}}{I_{M,M}} \nonumber \eeq
Now, it is easy to see that in computing $I_M$ the expression under exponent is zero. This means that the normalization of $\psi$ at $t_M$ implies that 
\beq I_{M,M} = 1 \eeq
Therefore, we can rewrite the above as 
\beq f_{M+1} (\vec{x}_{M+1}), \cdots, f_N (\vec{x}_N) = \bigg(\prod_{k=M+1}^N A_k \bigg) \bigg(\frac{\alpha}{\pi}\bigg)^{\frac{3}{2} (N-M)} I_{M,N} = \bigg(\prod_{k=M+1}^N A_k \bigg) \bigg(\frac{\alpha}{\pi}\bigg)^{\frac{3}{2} (N-M)} \times \nonumber \eeq
\beq \times \int d^3 x \bigg\vert \int_{\vec{x}' (t_k)= \vec{x}} [{\cal D} x'] \psi (\vec{x}', t_M^+) \exp \bigg(-\frac{\alpha}{2} \sum_{j=M+1}^k \vert \vec{x} - \vec{x}_j \vert^2 + i \int_{t_M}^{t_k} dt' \; {\cal L} (\vec{x}', t') \bigg) \bigg\vert^2 \eeq
This can be rewritten as 
\beq f_{M+1} (\vec{x}_{M+1}), \cdots, f_N (\vec{x}_N) = \eeq
\beq = \bigg(\prod_{k=M+1}^N A_k \bigg) \bigg(\frac{\alpha}{\pi}\bigg)^{\frac{3}{2} (N-M)} \int d^3 x \bigg\vert \int_{\vec{x}'(t_k) = \vec{x}} [{\cal D} x'] \psi (\vec{x}', t_M^+) e^{iS_{eff} (x')} \bigg\vert \nonumber \eeq
where the \emph{effective action} $S_{eff}$ is defined as
\beq S_{eff} (x') = \frac{i\alpha}{2} \sum_{j=M+1}^N \vert \vec{x} - \vec{x}_j \vert^2 + \int_{t_M}^{t_N} dt' \; {\cal L} (\vec{x}', t') \label{EffectiveAction} \eeq
Now, let us assume that $M-N$ is very large and on the large scale the ''hits'' fall into a continuous curve, 
\beq \vec{x}_k = \vec{\gamma}(t_k) \eeq
Furthermore, we will also assume that, on average, 
\beq t_{k+1} - t_k \approx \delta t \eeq
This will allow us to rewrite Eq \ref{EffectiveAction} as 
\beq S_{eff} \approx \int_{t_M}^{t_N} dt' \bigg({\cal L}(\vec{x}', t') + \frac{i \alpha}{2 \delta t} \vert \vec{x}' - \vec{\gamma} (t') \vert \bigg) \eeq
Thus we obtain an effective Lagrangian 
\beq {\cal L}_{eff} (\vec{x}', t') = {\cal L}(\vec{x}', t') + \frac{i \alpha}{2 \delta t} \vert \vec{x}' - \vec{\gamma} (t') \vert^2 \label{EffectiveLagrangian} \eeq
If we now define
\beq \epsilon_{eff} = \frac{\alpha}{\delta t} \eeq
we can rewrite the above as 
\beq {\cal L}_{eff} (\vec{x}', t') = {\cal L}(\vec{x}', t') + \frac{i \epsilon_{eff}}{2} \vert \vec{x}' - \vec{\gamma} (t') \vert^2 \label{EffectiveLagrangian} \eeq
The above $\epsilon_{eff}$ will later translate into the $i \epsilon$ term in QFT propagator. We can interpret $\vec{\gamma} (t')$ as a ''classical'' trajectory of the particle while $\vec{x}' (t') - \vec{\gamma} (t')$ as ''quantum fluctuation''. Thus, the above effective Lagrangian is used specifically for computing of probability that trajectory $\vec{\gamma} (t)$ takes place. 

\subsection*{Conclusion}

In this paper we have shown that GRW model on the smaller time scales might lead to Mensky's and Kent's model on larger time scales. In the Mensky's book, he have stated that his model might, indeed, emerge out of other quantum mechanical models. However, he has been emphasizing the quantum decoherence models, that rely on the separation between the "system" and "environment". Since both the "system" and "environment" consist of particles, those models are empirical rather than fundamental. On the other hand, in this paper we have replaced them with GRW model where all of the particles are being part of the "system" while the role of the environment is played by "hits" without any matter source. Thus, GRW model can be viewed as fundamental, and we have shown explicitly that it, too, leads to the weighted path integral. 

Kent's proposal (\cite{Kent}) takes up slightly different philosophy than Mensky's (\cite{Mensky1} and \cite{Mensky2}) in a sense that Kent tries to view weighted path integral as fundamental rather than empirical. This is both good and bad. On a good side, Kent's measurement outcome takes the form of spacetime trajectory as opposed to a state on a spacelike hypersurface, which makes it inherently more relativistic. On a bad side, it contradicts our intuitive notion of causality since the outcomes of past and future measurements are really just different parts of one single outcome, which means that the former does not cause the latter. Since this leads to a controversy with arguments on both sides, it is important to come up with models that accommodate both points of view. As far as this paper is concerned, we have attempted to accommodate the preference of causality over relativity, and shown how Kent's proposal still takes place in this context, it simply becomes emergent rather than fundamental. On the other hand, in \cite{Epsilon} we have taken the opposite poit of view  of relativity being preferred to causality. 

If we go within the context of QFT, which is the context in which \cite{Epsilon} was written, then the argument of this paper goes through if we replace $\psi (x, t)$ with $\psi (\phi, t)$ where $\phi$ is a scalar field defined over $t=const$ hypersurface. The specific prescription of how to do that is given at Section 3 of \cite{PsiOfPhi}. The notation $\psi (\phi, t)$, with $t$ present and $x$ absent, implies the violation of relativity; similarly, $\phi$, being defined over spacelike hypersurface, implies violation of relativity as well. \emph{Both of those statements are true even without GRW hits}. However, without the measurement, the violation of relativity "hides itself" due to the Lorentz covariant Lagrangian. In other words, there is an isomorphism between non-relativistic quantities defined in one frame and non-relativistic ones in another frame, and said isomorphism doesn't allow us to see just what frame happens to be preferred in the universe we live in, even though there is one! At the same time, if we do introduce GRW hits, then the fact that they are discrete rather than continuous breaks down the isomorphism which means that the direction of $t$-axis becomes physically relevent. On the other hand, Kent's model, which is continuous, is perfectly compatible with the isomorphism which is why it preserves Lorentz covariance. 

Nevertheless, if we try to make sense of mass density, we might still encounter problems: for example, if we are to identify it with energy-momentum tensor $T_{\mu \nu}$, such tensor would be tied together to conservation laws via Noether's theorem; since GRW model predicts increase of total energy, we have to rely on the fact that said increase occurs only during the ``hits", which means that energy is conserved during time intervals between them. Such statement, however, would no longer be true if we talk about Kent's continuous measurement model. So what we would have to do is to restrict ``correct" data to a hypersurface and then obtain ``wrong" data away from that hypersurface by extrapolating what ``would" happen ``if" energy-momentum was conserved. Now, since we need hypersurface to obtain pretend-conservation, and we need pretend-conservation in order to define $T_{\mu \nu}$, this means that $T_{\mu \nu}$ is a function of the hypersurface; in other words, it is not Lorentz invariant. Similarly, if instead of using $T_{\mu \nu}$ we were to use particle numbers, then we would run into well known fact that in curved spacetime QFT the number of particles is again frame dependent, which would result in frame dependent mass density as well. More detailed discussion of hypersurface dependence of $T_{\mu \nu}$ as well as possible ways of addressing it, is given in \cite{Referree5}

It should be pointed out that Tumulka made an attempt to reconcile GRW with relativity (see \cite{GRWTumulka}). His idea is that, while he still has a sequence of hypersurfaces on which to define quantum states, said hypersurfaces are defined in Lorentz covariant way. For example, instead of simply selecting $t=t_k$ hypersurface, he would rather select $(x^{\mu}-x_k^{\mu})(x_{\mu} - x_{k \mu}) = \tau_0^2$. From my point of view, this is still somewhat non-relativistic. After all, one could imagine that $\tau_0$ is of the size of a million light years and we are only a hundred light years away from said hyersurface. If we imagine that said hypersurface creates classical field around itself, then we can utilize the direction of such field in order to write down non-relativistic theory, such as Newton's second law, in Lorentz covariant way. So, from this point of view, Kent's proposal is the most relativistic of the three, mine is the most non-relativistic, and Tumulka's is somewhere in between (if I were to use simplest possible choice of hypersurfaces in "translating" the QM version of my model into QFT case). In any case, the argument that Kent's model can emerge from non-relativistic GRW can probably be modified in such a way that would show that Tumulka's GRW leads to emergence of Kent's model as well, but the explicit verification of this conjecture is beyond the scope of this paper. 

Let us go back to comparison of resent work and \cite{Epsilon}. In a nutshell, as far as \cite{Epsilon} is concerned, Kent's proposal arises naturally if one were to take $i \epsilon$ in the two point function
\beq \int d^4 p \frac{e^{i p \cdot x}}{p^2-m^2-i \epsilon} \eeq
literally and set $\epsilon$ to be finite. The finite value of $\epsilon$ determines the width of the weight function and, therefore, the scale where quantum ends and classical begins; thus, the assumption that $\epsilon$ is infinitesimal is equivalent to classicality arising only on infinitely large scales. The fact that $\epsilon$ is fundamental (after all, it is needed in order to avoid poles) implies that Kent proposal is, as well. That is why, on the first glance, the point of view of \cite{Epsilon} and the point of view of current paper are diametrically opposite when it comes to the question whether Kent is fundamental or emergent. 

However, one might take a different view point and reconcile the two papers. The way this can be done is by saying that yes Kent is still emergent, \emph{and so is $\epsilon$}. If we discretize all relevant parameters (space, time, energy, momentum, and so forth), then one would not need $\epsilon$ in order to avoid poles: if a straight line is replaced by sequence of points then those points will miss the poll, with absolute certainty. Discretization, by the way, is not such a big sacrifice since, in QFT case, discretization is needed anyway in order to avoid integration over uncountably many degrees of freedom, which can't be rigorously defined. However, despite the fact that two point function will end up well defined, it won't be the one we want: it will be a superposition of forward and backward moving ones, instead of just one of the two. But then, as a consequence of GRW "hits", on a large timescales the positive value of $\epsilon$ will emerge, resulting with only forward moving two point function "surviving". In other words, between two subsequent hits, the time reversal symmetry is satisfied. But then each given hit weakens the backward moving signal more than it weakens forward moving one. On the scale of one or two hits, each weakening is negligible and, therefore, inconsequential; but on the scale of lots of hits we end up "passing" forward moving signal and "stopping" backward moving one. Following this logic, the one and only source of time asymmetry is the asymmetry of GRW hit itself \cite{SelfReference}. This might lead to some ideas of exploring "more time symmetric hits" and, therefore, having both forward and backward propagators as well as phenomena such as post selection and so forth; but this is beyond the scope of this paper.


\begin{thebibliography}{77}

\bibitem{Epsilon} Link between quantum measurement and the i$\epsilon$  term in the quantum field theory propagator, Roman Sverdlov and Luca Bombelli. Phys. Rev. D 90, 125020 (2014) – Published 22 December 2014

\bibitem{Mensky1} M.B. Mensky, “Quantum continuous measurements, dynamical role of information and restricted path integrals”, in Proceedings TH2002 (International Conference on Theoretical Physics) Supplement, Birkh¨auser 2003, and arXiv:quant-ph/0212112.

\bibitem{Mensky2} M.B. Mensky Quantum Measurement and Decoherence Kluwer Academic Publishers 2000. 

\bibitem{Kent} A. Kent “Path integrals and reality” arXiv:1305.6565.

\bibitem{GRW1} G.C. Ghirardi, A. Rimini and T. Weber, “A model for a uniﬁed quantum description of macroscopic and microscopic systems”, in Quantum Probability and Applications, L. Accardi et al. (eds), Springer, Berlin, 1985. 

\bibitem{GRW2} G.C. Ghirardi, A. Rimini and T. Weber, “Uniﬁed dynamics for microscopic and macroscopic systems”, Phys. Rev. D 34, 470 (1986). 

\bibitem{GRWTumulka} Roderich Tumulka, "Collapse and Relativity", arXiv:quant-ph/0602208

\bibitem{PsiOfPhi} Roman Sverdlov "Expressing QFT in terms of QM with single extra dimension and classical hidden variable field", arXiv:1309.3287 

\bibitem{SelfReference} This argument has also been made in sec 4 of \cite{Epsilon}, where we were referring to the previous version of the current paper (arXiv:1305.7516, version 1). I chose to reiterate it in this paper since this is really the bridge between the two papers, equally relevent to both of them. 

\bibitem{Referree1} Describing the Macroscopic World: Closing the Circle within the Dynamical Reduction Program, Found. Phys, 25, 5 (1995), G.C. Ghirardi, R. Grassi, F. Benatti. This is the basic paper in which the interpretation has been presented and discussed for the first time.

\bibitem{Referree5} 5. Matter Density and Relativistic Models of Wave Function Collapse, Daniel Bedingham, Detlef Dürr, GianCarlo Ghirardi, Sheldon Goldstein, and Nino Zanghì.
Journal of Statistical Physics 154: 623-631 (2014)


\end{thebibliography}
\end{document}